\documentclass{article}\usepackage[]{graphicx}\usepackage[]{color}
\makeatletter
\def\maxwidth{ %
  \ifdim\Gin@nat@width>\linewidth
    \linewidth
  \else
    \Gin@nat@width
  \fi
}
\makeatother

\definecolor{fgcolor}{rgb}{0.345, 0.345, 0.345}

\usepackage{framed}
\makeatletter
 {\par\unskip\endMakeFramed%
 \at@end@of@kframe}
\makeatother

\definecolor{shadecolor}{rgb}{.97, .97, .97}
\definecolor{messagecolor}{rgb}{0, 0, 0}
\definecolor{warningcolor}{rgb}{1, 0, 1}
\definecolor{errorcolor}{rgb}{1, 0, 0}
\newenvironment{knitrout}{}{} 

\usepackage{alltt}

\usepackage[utf8]{inputenc}
\usepackage[T1]{fontenc}

\usepackage{amsmath}
\usepackage{amssymb}
\usepackage{amsfonts}
\usepackage{graphicx}
\usepackage{xspace}
\usepackage{listings}
\usepackage{url}
\usepackage{algorithm}
\usepackage{algorithmicx}
\usepackage[noend]{algpseudocode}
\usepackage{subfigure}

\usepackage{multirow}
\usepackage{booktabs}
\usepackage{tikz}
\usepackage{pifont}
\usepackage{standalone}
\usepackage{microtype}
\usepackage{tabularx}
\usepackage{float}
\usepackage{placeins}
\usepackage[english]{babel}
\usepackage[autolanguage]{numprint}
\usepackage[binary-units]{siunitx}
\usepackage{todonotes}
\usepackage{hyperref}
\usepackage[title]{appendix}
\usepackage[letterpaper,margin=3cm]{geometry} 
\usepackage{libertine}

\presetkeys%
    {todonotes}%
    {inline,backgroundcolor=yellow}{}

\newcommand{\append}{Appendix\xspace}
\newcommand{\Sec}{Section\xspace}

\newcommand{\eg}{e.g.\xspace}

\newcommand{\etal}{et~al.\@\xspace}

\newcommand{\cf}{cf.\@\xspace}

\newcommand{\reppar}{REPPAR\xspace}

\newcommand{\europar}{Euro-Par\xspace}

\newcommand{\ques}[2]{Q2.#1.#2\xspace}

\newcommand{\runtime}{run-time\xspace}

\date{\vspace{-5ex}}
\IfFileExists{upquote.sty}{\usepackage{upquote}}{}
\begin{document}

\title{A Survey on Reproducibility in Parallel Computing}

\author{
Sascha Hunold\\[1ex]
Vienna University of Technology\\Faculty of
    Informatics\\ 
    Research Group for Parallel Computing\\
    Favoritenstra\ss{}e 16/184-5\\1040 Vienna, Austria\protect\\
    \texttt{hunold@par.tuwien.ac.at}
}

\maketitle

\begin{abstract}
  We summarize the results of a survey on reproducibility in parallel
  computing, which was conducted during the \europar conference in
  August 2015. The survey form was handed out to all participants of
  the conference and the workshops. The questionnaire, which
  specifically targeted the parallel computing community, contained
  questions in four different categories: general questions on
  reproducibility, the current state of reproducibility, the
  reproducibility of the participants' own papers, and questions about
  the participants' familiarity with tools, software, or open-source
  software licenses used for reproducible research.
\end{abstract}

\section{Introduction}

Conducting sound and reproducible experiments in parallel computing is
not easy, as hardware and software architectures of current parallel
computers are most often very complex. This high complexity makes it
difficult and often impossible for scientists to model such systems
mathematically. Thus, scientists often rely on experiments to study
new parallel algorithms, different software solutions (\eg, operating
systems), or novel hardware architectures. The situation in parallel
computing is made even more difficult as parallel systems are in a
constant state of flux, \eg, the total core count is rapidly growing
and many programming paradigms for parallel machines have emerged  and
are actively being used.

We established the first edition of the International Workshop on
Reproducibility in Parallel Computing
(\reppar\footnote{\url{http://reppar.org/}}) in conjunction with the
\europar conference in 2014.  The workshop is concerned with
experimental practices in parallel computing research.  It should be a
forum for discussing and exchanging ideas to improve reproducibility
matters in our research domain. We solicit research papers and
experience reports on a number of relevant topics, particularly:
methods for analysis and visualization of experimental data, best
practice recommendations, results of attempts to replicate previously
published experiments, and tools for experimental computational
sciences. Some examples of the latter include workflow management
systems, experimental test-beds, and systems for archiving and
querying large data files.

In 2015, the \reppar workshop was hosted for the second time in
conjunction with the \europar conference. This year we wanted to spark
a fruitful discussion by conducting a survey on reproducible research
and by evaluating the results directly during the workshop.
In the present paper, we will take a closer look at the results of the
survey and discuss some of the 
findings.

After summarizing related work in \Sec~\ref{sec:rel_work}, we explain
the context of the survey in \Sec~\ref{sec:survey}.
\Sec~\ref{sec:survey_results} presents the survey results, and we draw
conclusions in \Sec~\ref{sec:conclusions}.

\section{Related Work}
\label{sec:rel_work}

Improving the reproducibility of results that get published in today's
scientific journals is one of the big challenges of the current
research landscape, not only because the problem has lately been
brought into the spotlight by journals like Science or Nature
(\cf~\cite{Nature2013,Buck_Science}). Thus, many researchers across
disciplines are trying to tackle the problem of the irreproducibility
of scientific findings.

From a computer-science standpoint, we are foremost interested in the
state of reproducibility of computational results. The reproduction of
scientific findings in computational sciences has other challenges
than, say, medicine, as here we study abstract objects, \eg, a
computer program or an algorithm (rather than the human
body). Questions that arise in this context are, for example, how to
share source code (technically) or which license to apply to a piece
of software?  Stodden, Leisch, and Peng addressed these issues and
published a collection of articles, in which several solutions to the
dilemma are proposed~\cite{stodden:implementing}.

Here, we are not only interested in computational sciences, but
specifically in parallel and distributed computing, where we are
facing additional challenges in terms of
reproducibility~\cite{HunoldT13}. For example, in the high performance
computing community, scientists are primarily interested in optimizing
performance, \eg, trying to minimize the \runtime or to maximize the
throughput of a system. Thus, a reproducible analysis does not only
need to be able to solve the computational problem with the same
outcome, but also in the same---or at least comparable---time as shown
in the original paper.

Therefore, we conducted a survey on reproducible research among the
Euro-Par participants to gain insights about how the reproducibility
problem is perceived in our community. In the USA, several initiatives
or workshops exist that address the reproducibility problem for
large-scale computing. One example is the XSEDE workshop on
reproducibility~\cite{JamesWS14}.

Several surveys have been undertaken in the broader context of
reproducible research. For us, most related to our work are the survey
of Stodden~\cite{stodden:reproducibility} and the survey of
Prabhu~\etal \cite{Prabhu:2011}.  Stodden's survey sheds light on the
incentives for scientists to share or not to share their work (code or
data). The survey by Prabhu~\etal is more concerned with best
practices in computational sciences, for example, the authors try to
answer questions like ``do scientists know about parallelization
techniques for speeding up their applications'' or ``what languages do
scientists use for their daily compute tasks''. The survey results by
Prabhu~\etal reveal that ``[s]cientists should release code to their
peers'' in order to ``allow other scientists to reproduce prior
work''~\cite{Prabhu:2011}.

\section{Context of the Survey}
\label{sec:survey}

To conduct our survey on reproducibility in parallel computing, we
prepared a questionnaire containing 24~questions (\cf
\append~\ref{sec:questionnaire}), which we grouped into four different
categories (\cf \Sec~\ref{sec:general}--\Sec~\ref{sec:tools}). All
participants of the \europar conference received one survey
flyer, 
which advertised and introduced the survey to them. The flyer
contained a unique token that enabled each participant to vote exactly
once. The survey was completely anonymous and since flyers (and their
tokens) were handed out in the order in which participants arrived at
the conference registration desk, the identity of the voters was
additionally protected. The survey was implemented using the
LimeSurvey software\footnote{\url{http://www.limesurvey.org/}}.  We
printed 300 survey flyers, each containing one token, and handed out
one flyer to each of the 232 participants of the \europar
conference. Unfortunately, only 31 persons~(13\%) completed the online questionnaire.

\newpage
\section{Survey Results}
\label{sec:survey_results}

Now, we present the survey results and comment on the outcome of
individual questions.

\subsection{General Questions on Reproducibility}
\label{sec:general}

The first question (\ques{1}{1}) directly asked whether the survey
participant is interested in reproducible research. Rather
surprisingly, the majority of the participants ($>90\%$) declared
their interest in reproducibility. Considering the fact that only 31
persons completed the survey, we conclude that most of the
participants also attended the \reppar workshop. Therefore, we should
keep in mind that our results are highly biased towards a small group
of people sharing similar interests.

We also assumed that only few scientists know what ``reproducible
research'' means and also what the difference is between the terms
``replicability'', ``repeatability'', and ``reproducibility''.  Our
assumption was based on the fact that many articles use different
definitions of ``reproducibility''. Since we had posed a vague
question, the poll results of questions \ques{1}{2} and \ques{1}{3}
were surprising. For example, only 13\% of the voters stated that they
do not know the difference between replicability, repeatability, and
reproducibility.

It is also noteworthy that all survey participants think that the
reproduction of already published results is worth another
publication. However, 65\% of them demand that articles reproducing
work of others need to contain new insights.


\subsubsection{Do you care (in general) about the reproducibility of scientific results (your own, others)?}

\begin{knitrout}
\definecolor{shadecolor}{rgb}{0.969, 0.969, 0.969}\color{fgcolor}

{\centering \includegraphics[width=.8\linewidth]{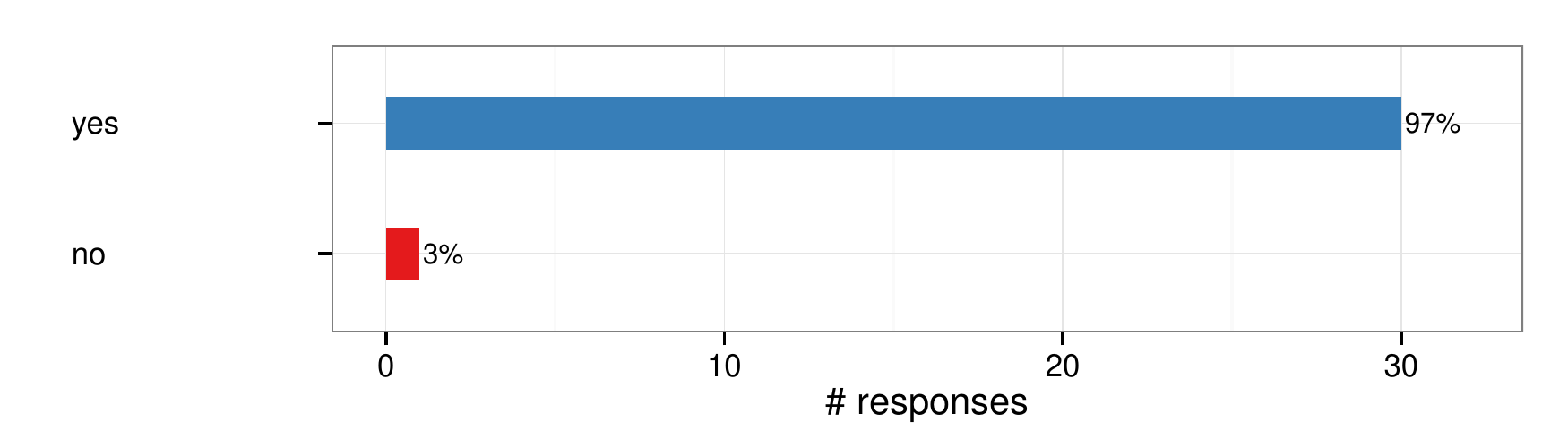} 

}

\end{knitrout}


\subsubsection{Do you know what people mean when speaking about "reproducible" results?}

\begin{knitrout}
\definecolor{shadecolor}{rgb}{0.969, 0.969, 0.969}\color{fgcolor}

{\centering \includegraphics[width=.8\linewidth]{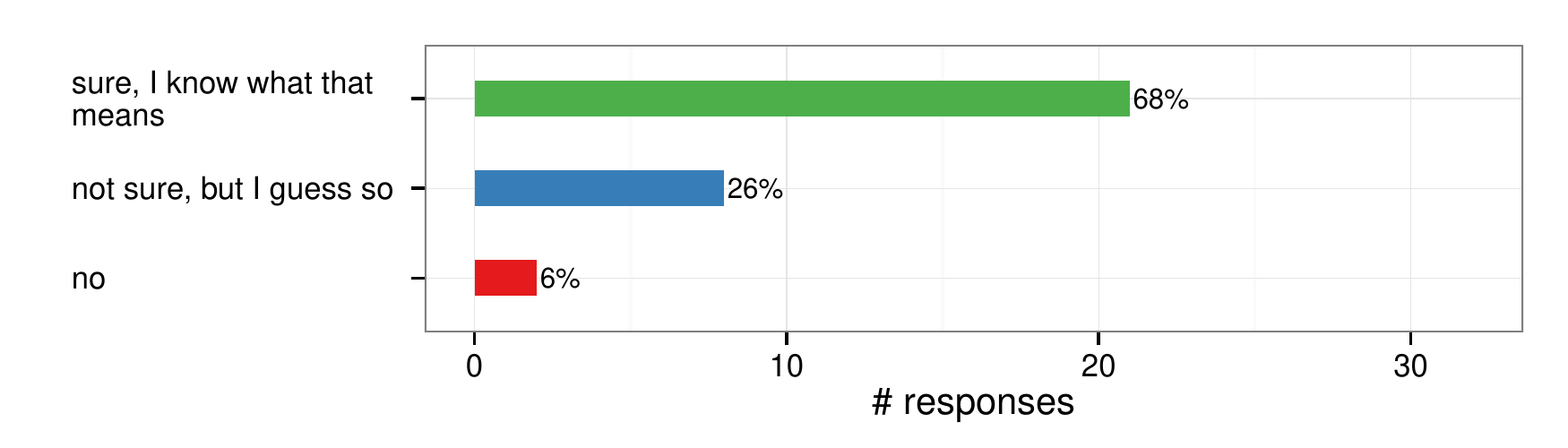} 

}

\end{knitrout}

\newpage

\subsubsection{Do you know the difference between replicability, repeatability, and reproducibility?}

\begin{knitrout}
\definecolor{shadecolor}{rgb}{0.969, 0.969, 0.969}\color{fgcolor}

{\centering \includegraphics[width=.8\linewidth]{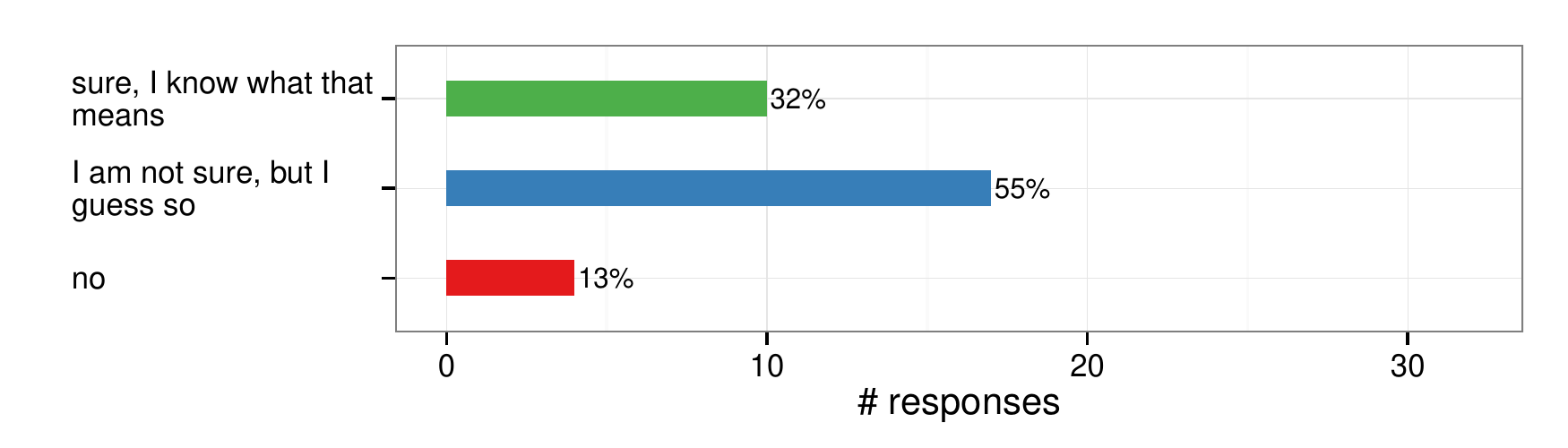} 

}

\end{knitrout}


\subsubsection{Do you think that the reproduction of already published results is worth another publication?}

\begin{knitrout}
\definecolor{shadecolor}{rgb}{0.969, 0.969, 0.969}\color{fgcolor}

{\centering \includegraphics[width=.8\linewidth]{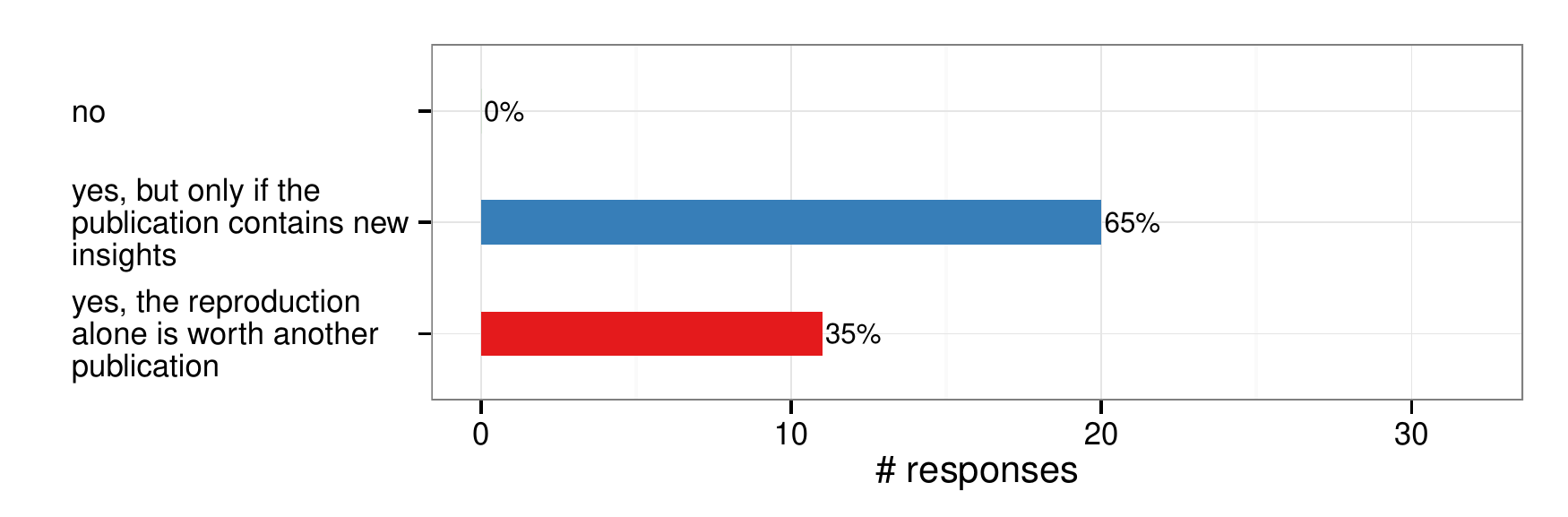} 

}

\end{knitrout}


\subsubsection{Have you tried to reproduce the results of others?}

\begin{knitrout}
\definecolor{shadecolor}{rgb}{0.969, 0.969, 0.969}\color{fgcolor}

{\centering \includegraphics[width=.8\linewidth]{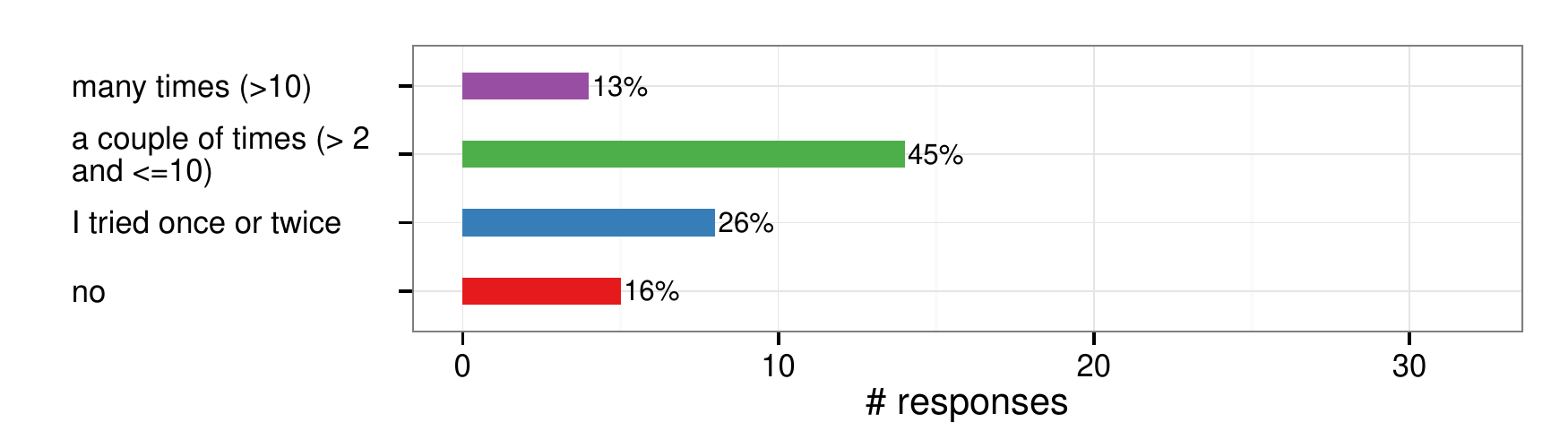} 

}

\end{knitrout}

\subsection{Current State of Reproducibility in Parallel Computing}
\label{sec:current_state}

With the second block of questions, we intended to learn more about
how scientists see the domain of parallel or high performance
computing (HPC) in terms of reproducibility.

The results of question \ques{2}{1} show a clear picture: almost all
participants think that the reproducibility of articles in our
research domain need to be improved. We again note that our results
are biased, as many of the survey participants also attended the
\reppar workshop on reproducible research.

It is also remarkable that only a small percentage (6\%) of the voters
believed that articles from top conferences such as PPoPP or IPDPS are
better reproducible than papers of other conferences (\ques{2}{3}).
On the contrary, many people (>50\% in sum) do not trust the
reproducibility of the results when they review scientific articles
(\ques{2}{4}).

\newpage

\subsubsection{Do you think the state of reproducibility for articles in our research domain (Parallel Computing/HPC) needs to be improved?}

\begin{knitrout}
\definecolor{shadecolor}{rgb}{0.969, 0.969, 0.969}\color{fgcolor}

{\centering \includegraphics[width=.8\linewidth]{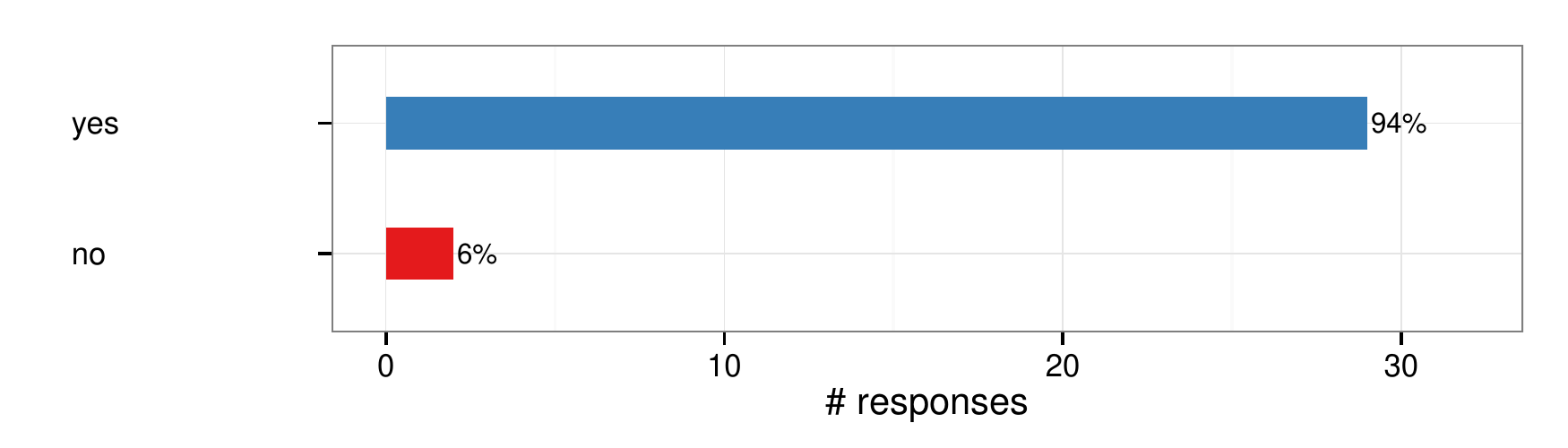} 

}

\end{knitrout}


\subsubsection{Do you think current research articles in the domain of Parallel Computing/HPC are reproducible by other independent researchers?}

\begin{knitrout}
\definecolor{shadecolor}{rgb}{0.969, 0.969, 0.969}\color{fgcolor}

{\centering \includegraphics[width=.8\linewidth]{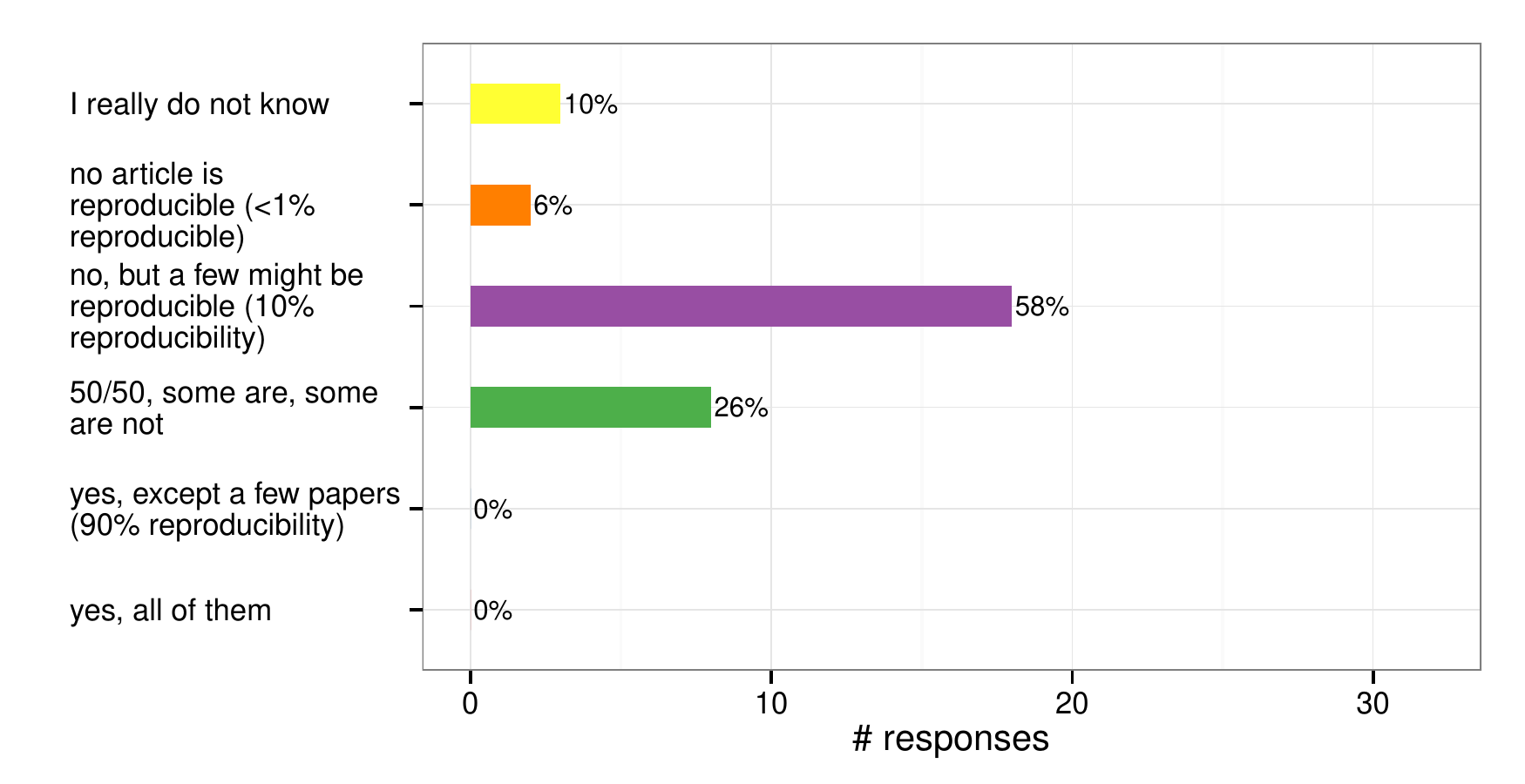} 

}

\end{knitrout}


\subsubsection{Do you think that results published in top conferences (e.g., PPoPP, IPDPS) are generally easier to reproduce than those published in lower-tier conferences in parallel computing (in the last 5 years)?}

\begin{knitrout}
\definecolor{shadecolor}{rgb}{0.969, 0.969, 0.969}\color{fgcolor}

{\centering \includegraphics[width=.8\linewidth]{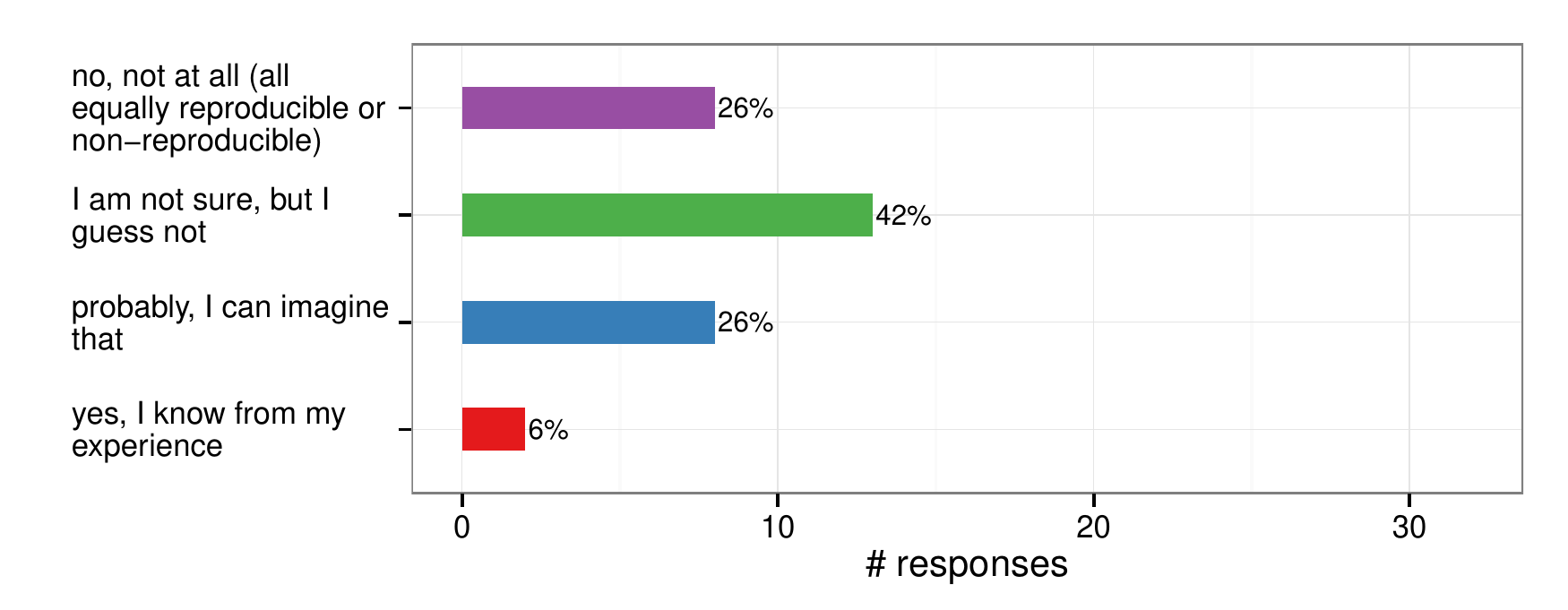} 

}

\end{knitrout}

\newpage

\subsubsection{How often do you question the reproducibility of results when you review other scientific articles?}

\begin{knitrout}
\definecolor{shadecolor}{rgb}{0.969, 0.969, 0.969}\color{fgcolor}

{\centering \includegraphics[width=.8\linewidth]{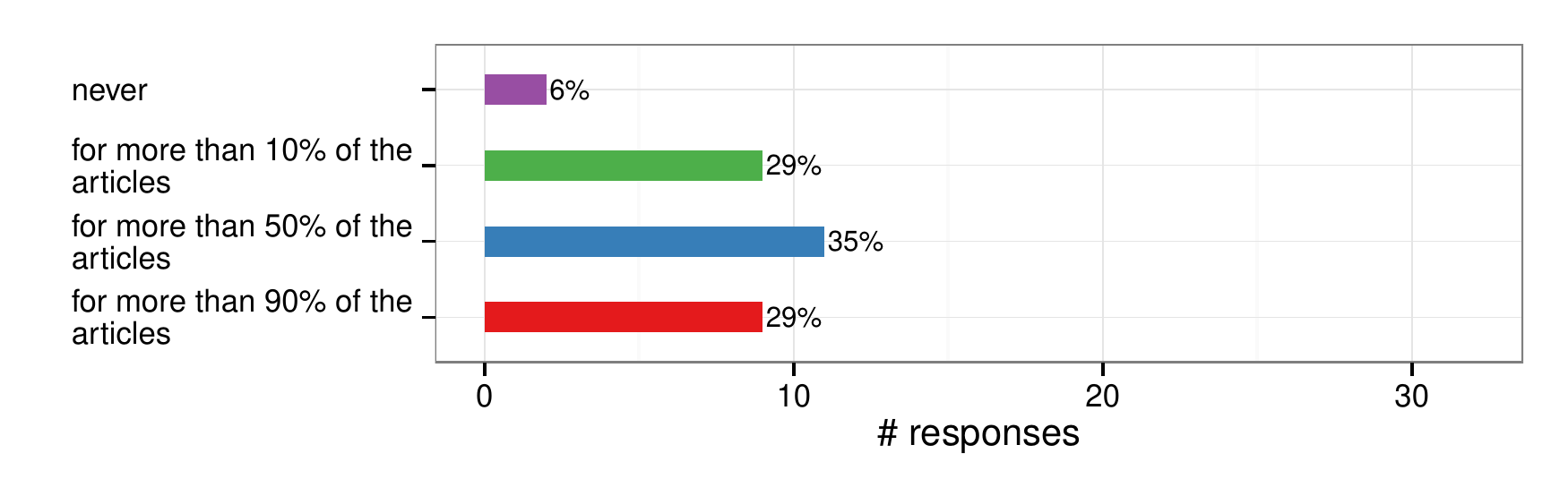} 

}

\end{knitrout}

\subsection{Reproducibility of Your Articles}
\label{sec:your_articles}

The third block of questions was concerned with what the participants
think about the reproducibility of their own articles.  The poll
results for question \ques{3}{1} show that a significant fraction of
the voters (19\%) believe that the results published in their articles
are reproducible by others. Surprisingly, only 3\% stated that they
know that their papers are not reproducible. We had expected a higher
percentage of people that would admit that their papers are hard to
reproduce, especially when taking into account that the poll was
anonymous.

90\% of the participants consider freely accessible HPC systems a
necessity for reproducible results. 

We also asked how scientists provide the source code, the raw
experimental data, and the data analysis procedures to others. Again,
it was surprising that a large percentage (23\%) of scientists stated
that they publish the source code along with their papers
(\ques{3}{2}). From our personal experience we had expected much less
(around 10\%).  The poll results also show that more than half of the
scientists use a public revision control system, such as GitHub, to
share their code (\ques{3}{4}).

However, when we look at the percentage of scientists that do not
provide the source code, the raw experimental data, or the data
analysis procedures, we can observe that the data analysis procedures
get shared less often compared to the other two. One explanation could
be that the data analysis procedures applied are very simple (\eg,
computing the arithmetic mean). Another explanation could be that
researchers simply do not give them a high priority, and perhaps do
not see the importance for others to have these procedures.

We also asked the survey participants about their main reasons for not
sharing code, data, or data analysis procedures (\ques{3}{8}).  Here,
no clear line can be drawn, as no answer was mentioned significantly
more often than others. Similarly, we did not obtain a clear picture
when asking the participants what they believe are the major obstacles
to reproduce their papers (\ques{3}{9}).



\subsubsection{Do you think the results (contribution) published in YOUR papers are reproducible by others (in the last 5 years)?}

\begin{knitrout}
\definecolor{shadecolor}{rgb}{0.969, 0.969, 0.969}\color{fgcolor}

{\centering \includegraphics[width=.8\linewidth]{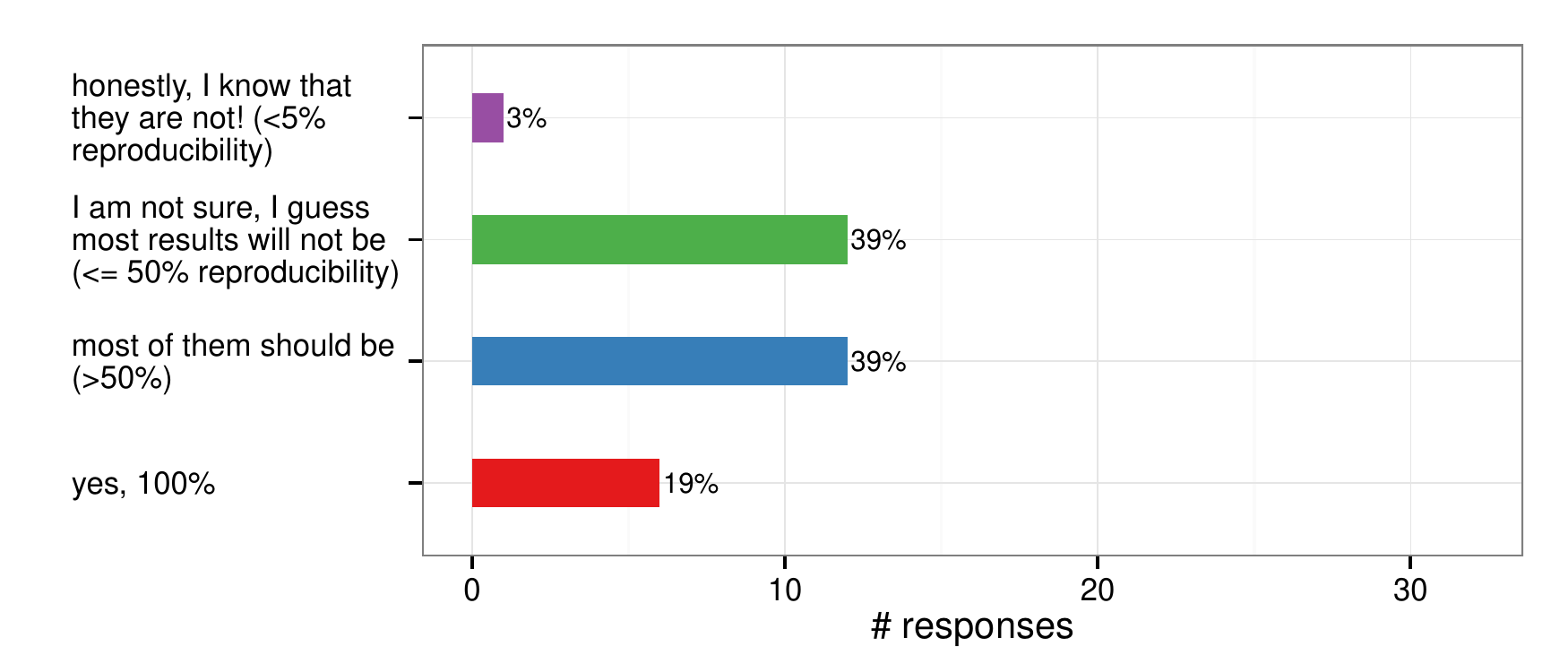} 

}

\end{knitrout}


\subsubsection{How often have you published the source code along with YOUR paper (in the last 5 years)?}

\begin{knitrout}
\definecolor{shadecolor}{rgb}{0.969, 0.969, 0.969}\color{fgcolor}

{\centering \includegraphics[width=.8\linewidth]{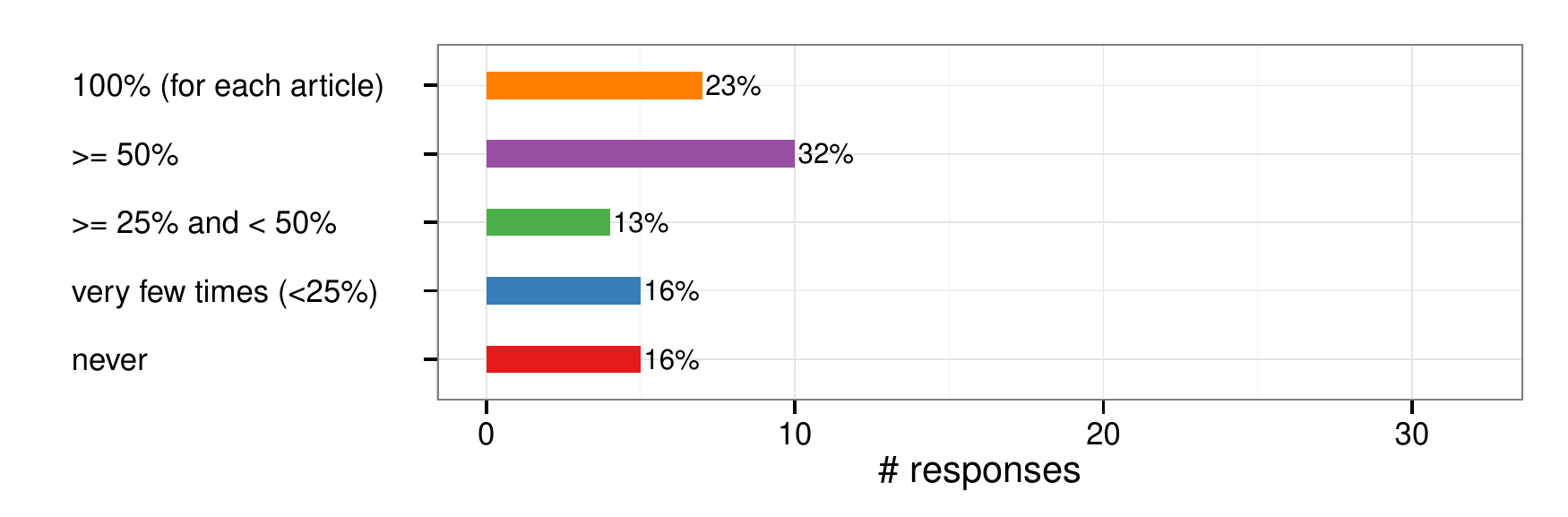} 

}

\end{knitrout}


\subsubsection{Do you consider freely accessible HPC systems a necessity for getting reproducible performance figures?}

\begin{knitrout}
\definecolor{shadecolor}{rgb}{0.969, 0.969, 0.969}\color{fgcolor}

{\centering \includegraphics[width=.8\linewidth]{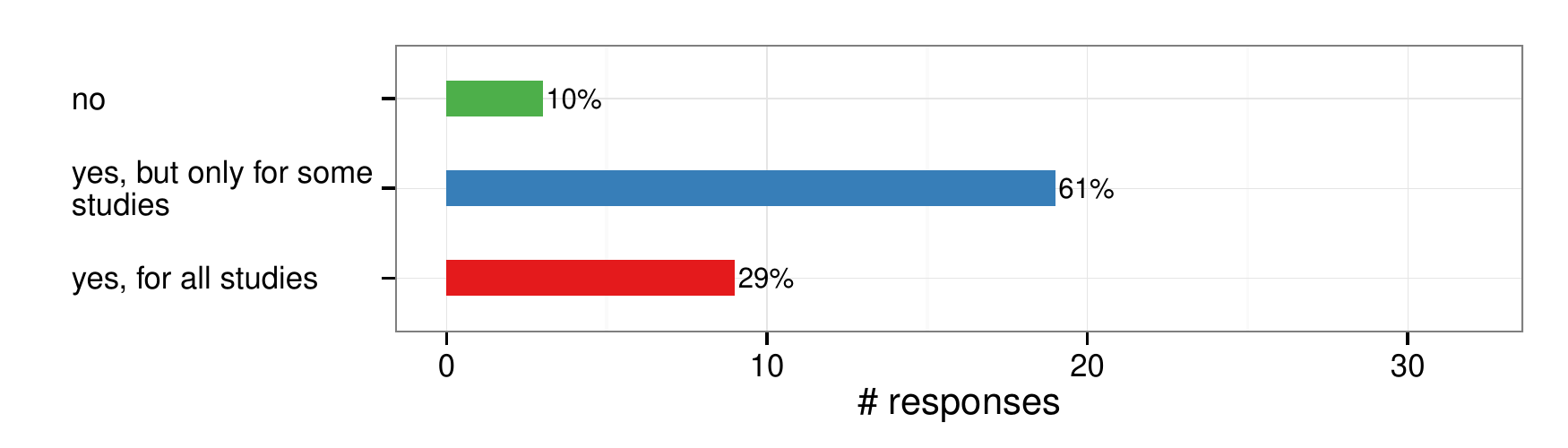} 

}

\end{knitrout}


\subsubsection{How do you provide "source code" for others?}

\begin{knitrout}
\definecolor{shadecolor}{rgb}{0.969, 0.969, 0.969}\color{fgcolor}

{\centering \includegraphics[width=.8\linewidth]{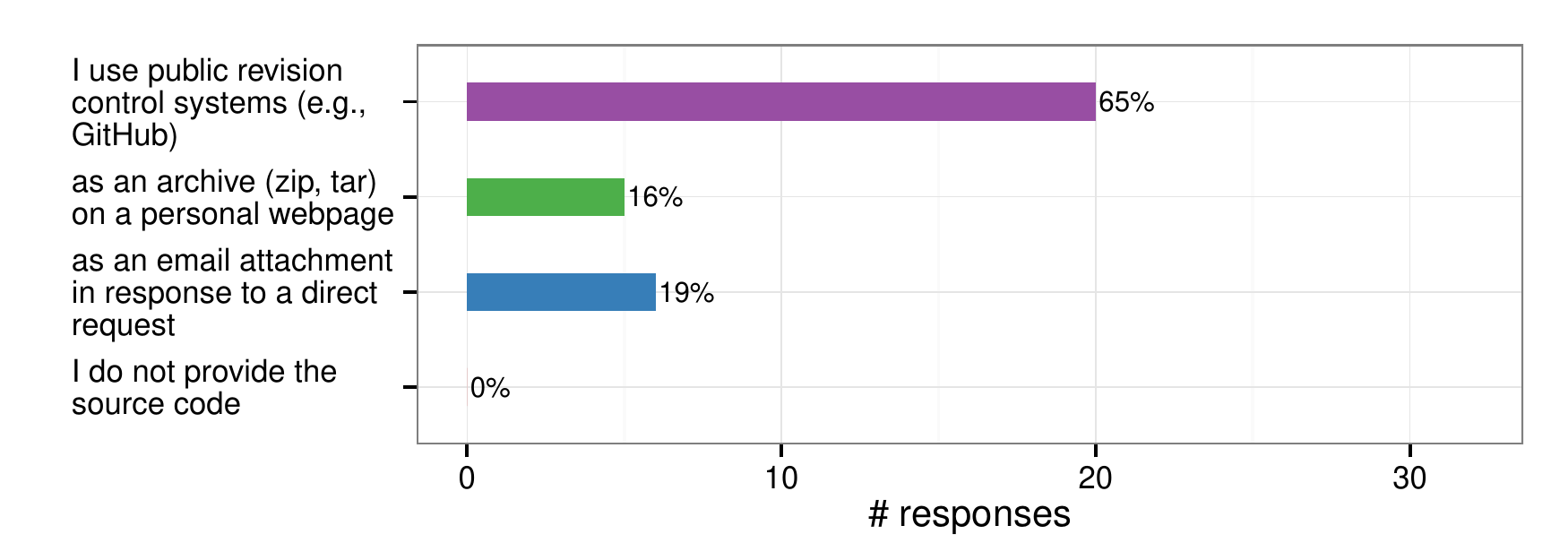} 

}

\end{knitrout}


\subsubsection{How do you provide the "raw data (of experiments)" for others?}

\begin{knitrout}
\definecolor{shadecolor}{rgb}{0.969, 0.969, 0.969}\color{fgcolor}

{\centering \includegraphics[width=.8\linewidth]{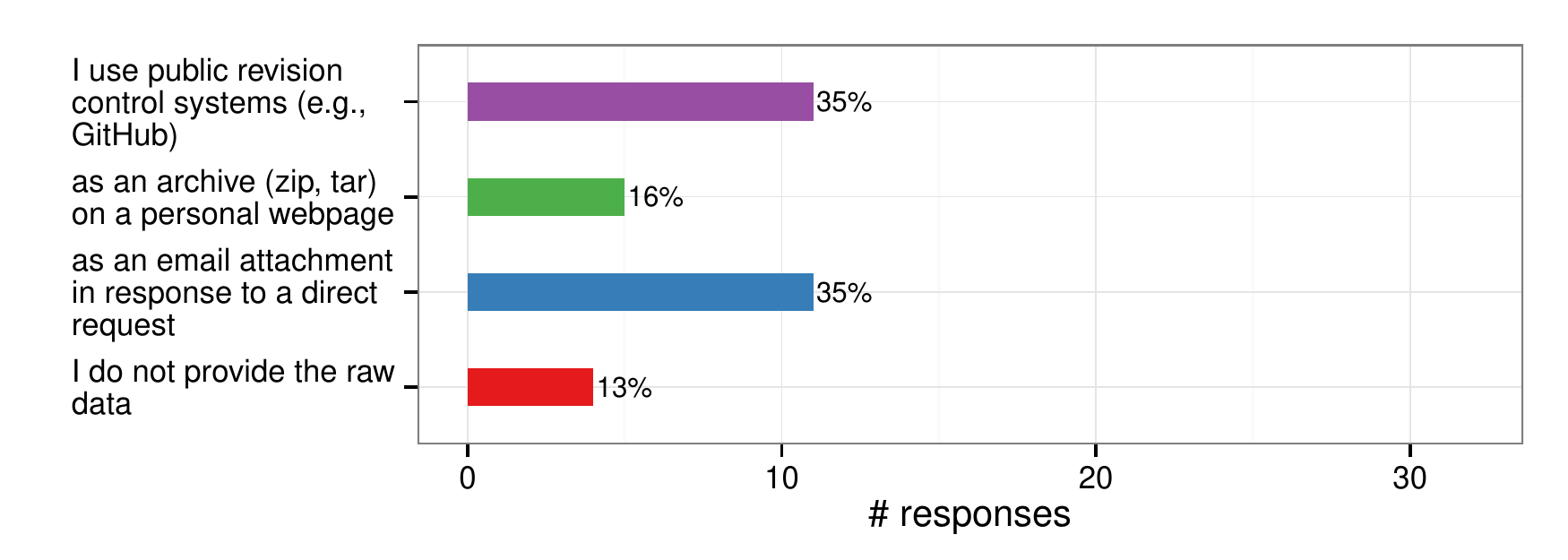} 

}

\end{knitrout}


\subsubsection{How do you provide the "data analysis procedure (R scripts, etc)" for others?}

\begin{knitrout}
\definecolor{shadecolor}{rgb}{0.969, 0.969, 0.969}\color{fgcolor}

{\centering \includegraphics[width=.8\linewidth]{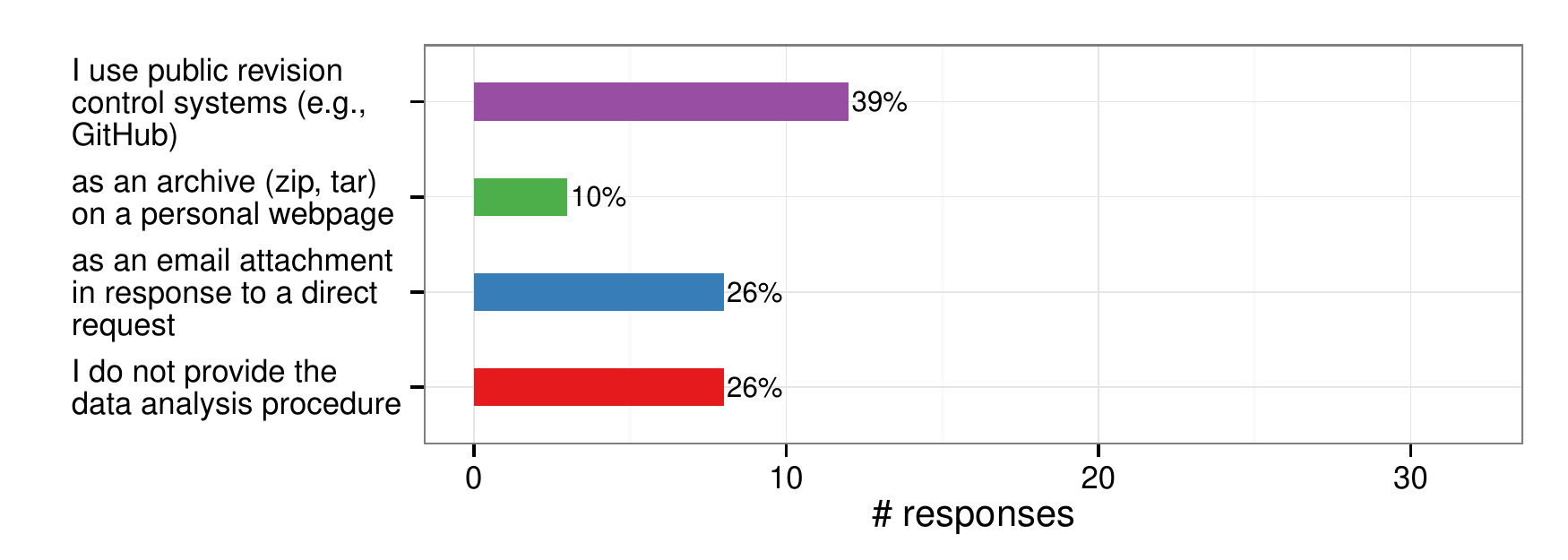} 

}

\end{knitrout}


\subsubsection{How do you document how to use your source code / data analysis scripts for others?}

\begin{knitrout}
\definecolor{shadecolor}{rgb}{0.969, 0.969, 0.969}\color{fgcolor}

{\centering \includegraphics[width=.8\linewidth]{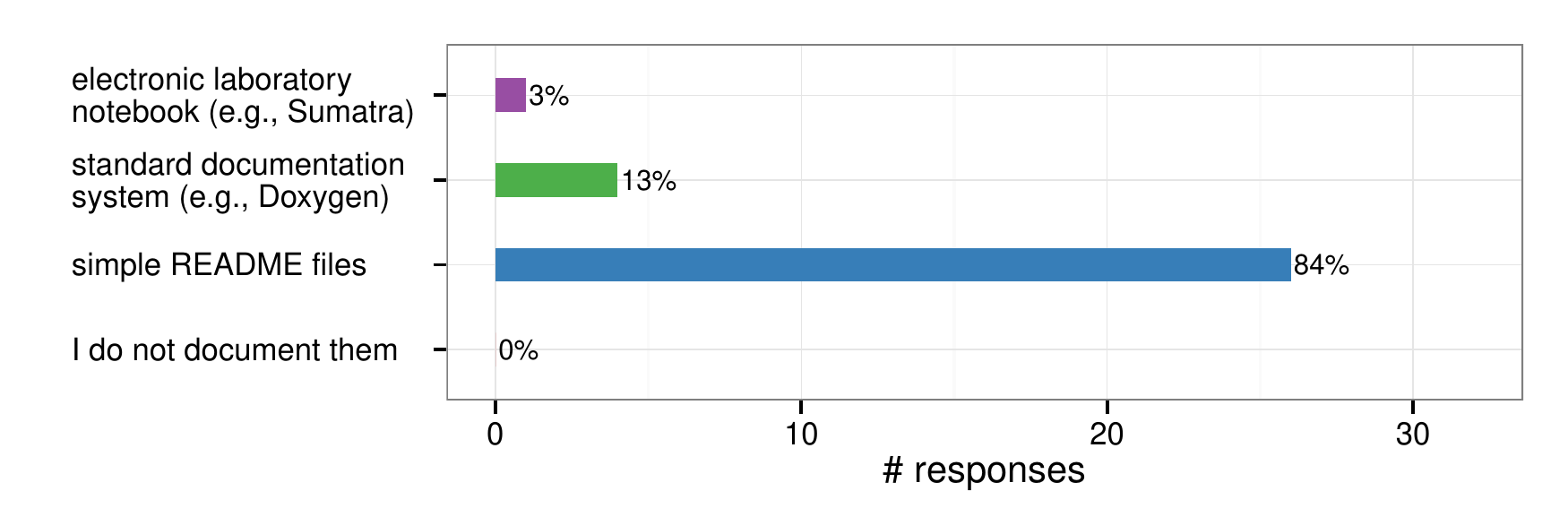} 

}

\end{knitrout}


\subsubsection{What are the main reasons for NOT making the source code/raw data/data analysis procedure available?}

\begin{knitrout}
\definecolor{shadecolor}{rgb}{0.969, 0.969, 0.969}\color{fgcolor}

{\centering \includegraphics[width=.8\linewidth]{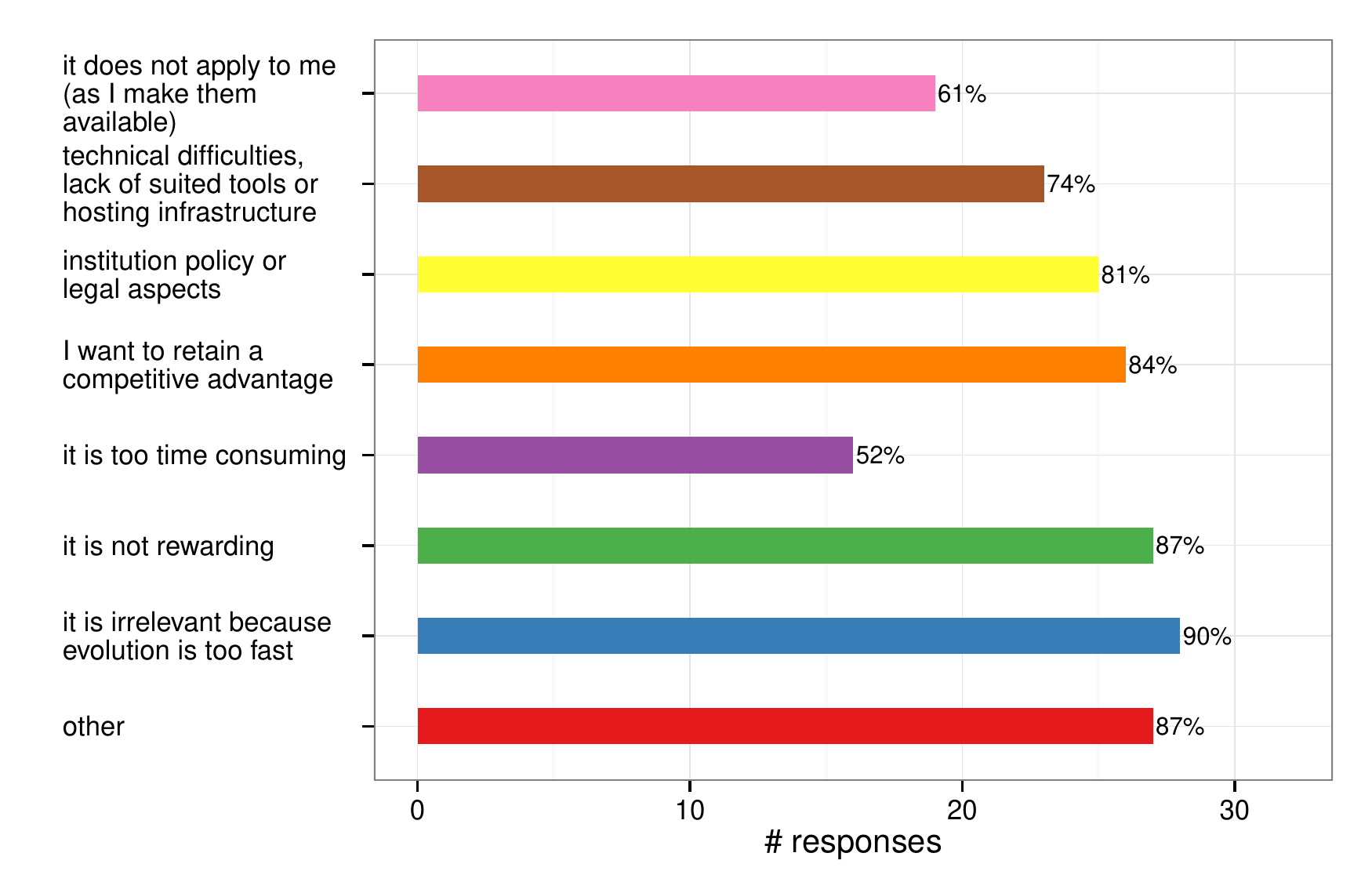} 

}

\end{knitrout}

\newpage

\subsubsection{What do you think will be the main difficulties/obstacles when other independent researchers try to reproduce YOUR experiments?}

\begin{knitrout}
\definecolor{shadecolor}{rgb}{0.969, 0.969, 0.969}\color{fgcolor}

{\centering \includegraphics[width=.8\linewidth]{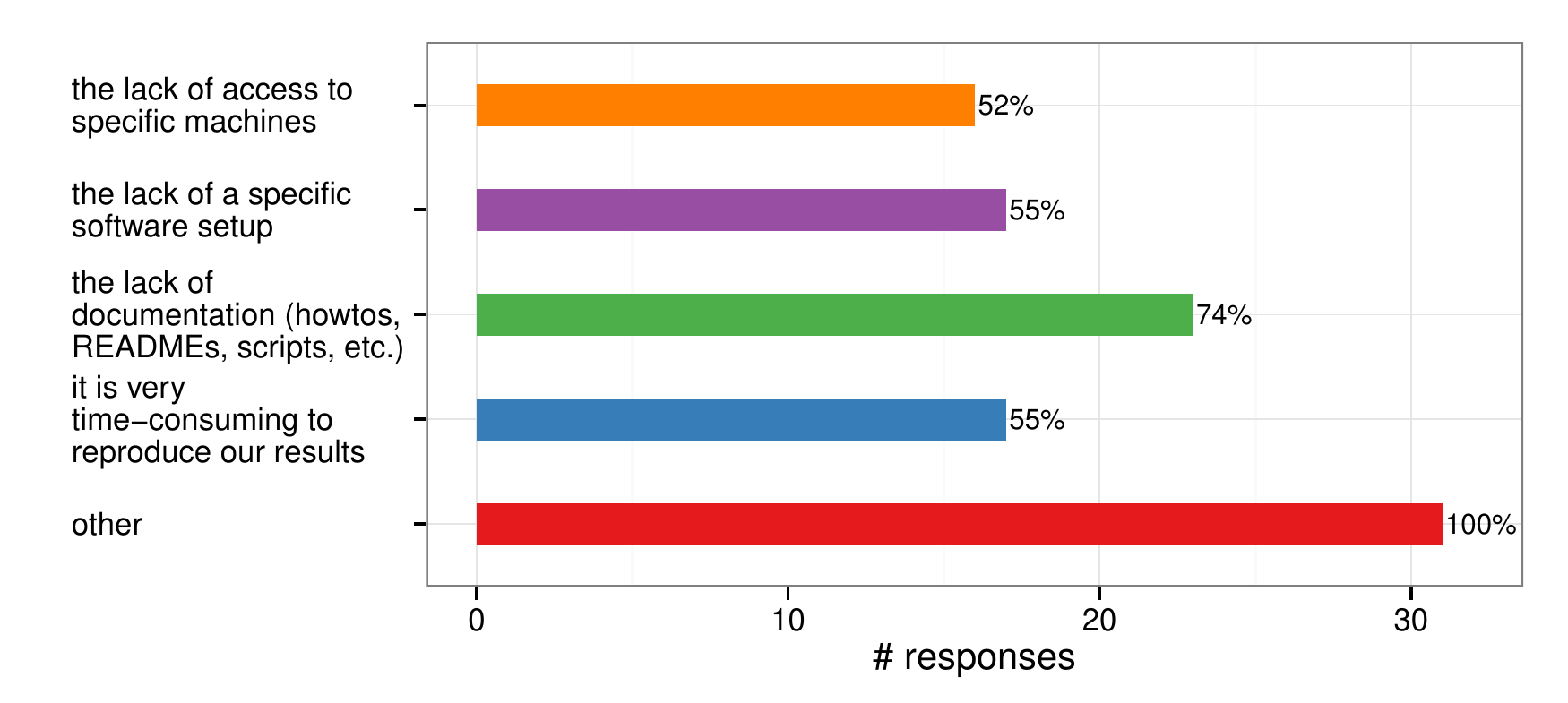} 

}

\end{knitrout}

\subsection{Tools, Software, and Licenses for Reproducible Research}
\label{sec:tools}

Last, we wanted to examine which software and licenses scientists use
for making their experiments reproducible.

In question \ques{4}{1}, we asked the participants whether they use
statistical software packages, such as R or SPSS, for performing data
analysis tasks. It turns out that only a third of the voters use such
tools on a regular basis. It is also remarkable that most of the
voters (71\% and 84\% respectively) had never used software for
literate programming (\eg, knitr or org-mode) nor tools for managing
or executing scientific workflows (\eg, VisTrails or Kepler).

Researchers often debate what open-source software license is the best
for their purposes. We therefore asked the question whether the
participants do know the license policy of their research institutions
(\ques{4}{6}). Only 19\% of the voters know this policy, whereas 26\%
stated that the institute has no explicit policy.


\subsubsection{Have you used statistical software packages (e.g., R, SAS, SPSS) for analyzing your experimental results?}

\begin{knitrout}
\definecolor{shadecolor}{rgb}{0.969, 0.969, 0.969}\color{fgcolor}

{\centering \includegraphics[width=.8\linewidth]{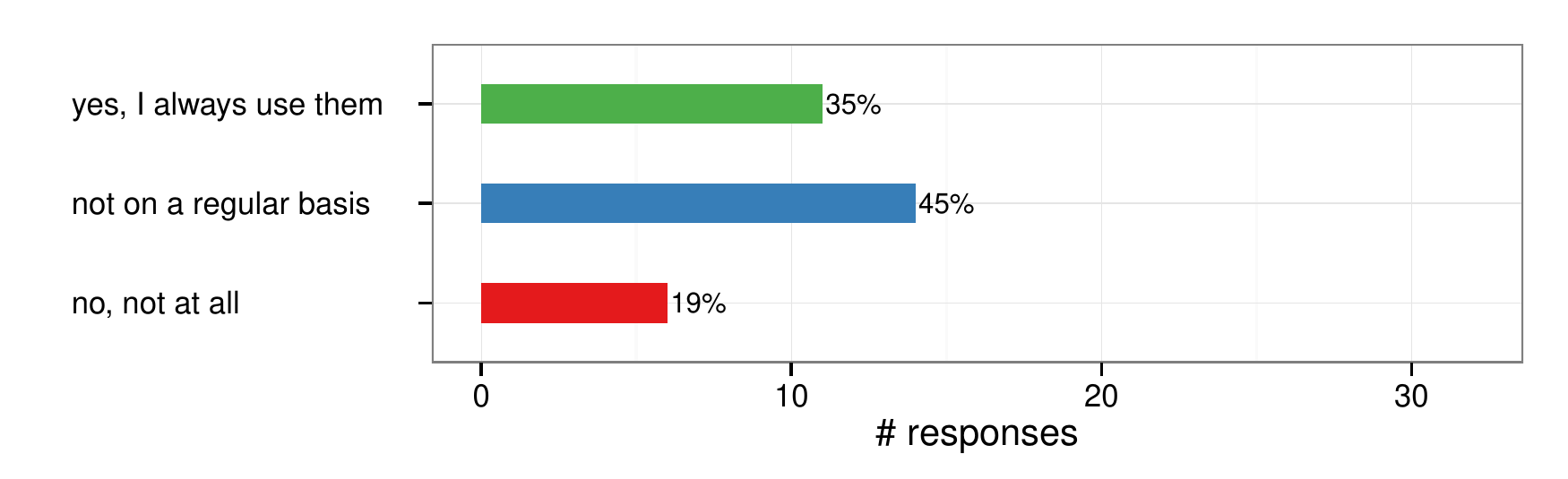} 

}

\end{knitrout}

\newpage

\subsubsection{How would you rate YOUR knowledge of the programming language "R"?}

\begin{knitrout}
\definecolor{shadecolor}{rgb}{0.969, 0.969, 0.969}\color{fgcolor}

{\centering \includegraphics[width=.8\linewidth]{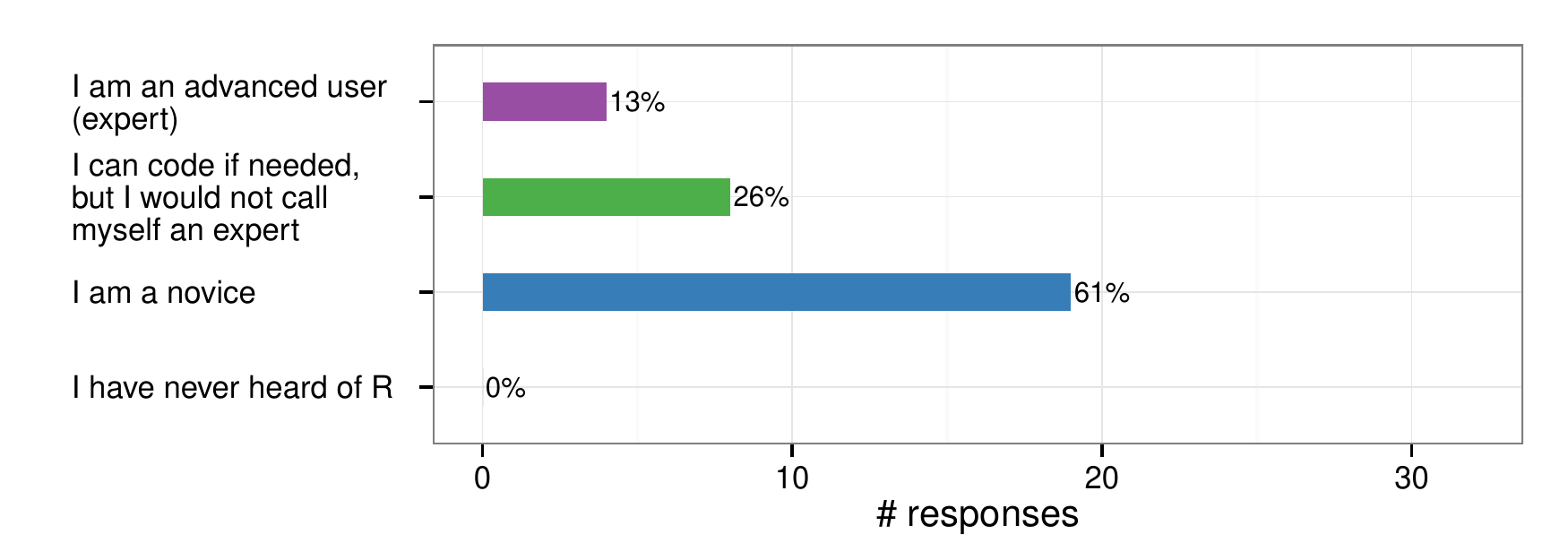} 

}

\end{knitrout}


\subsubsection{Do you use/have you used tools for literate programming (e.g., knitr, org-mode) for publishing articles?}

\begin{knitrout}
\definecolor{shadecolor}{rgb}{0.969, 0.969, 0.969}\color{fgcolor}

{\centering \includegraphics[width=.8\linewidth]{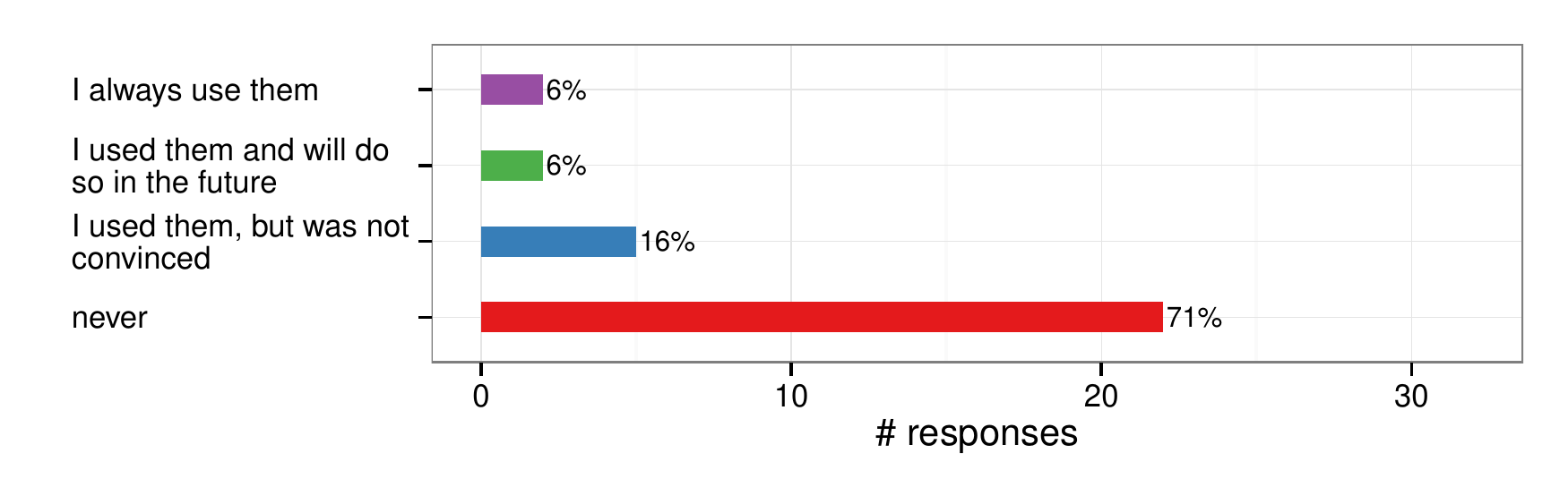} 

}

\end{knitrout}


\subsubsection{Do you have practical experiences with workflow tools to support reproducible research (e.g., VisTrails, Kepler, DataMill, etc.)?}

\begin{knitrout}
\definecolor{shadecolor}{rgb}{0.969, 0.969, 0.969}\color{fgcolor}

{\centering \includegraphics[width=.8\linewidth]{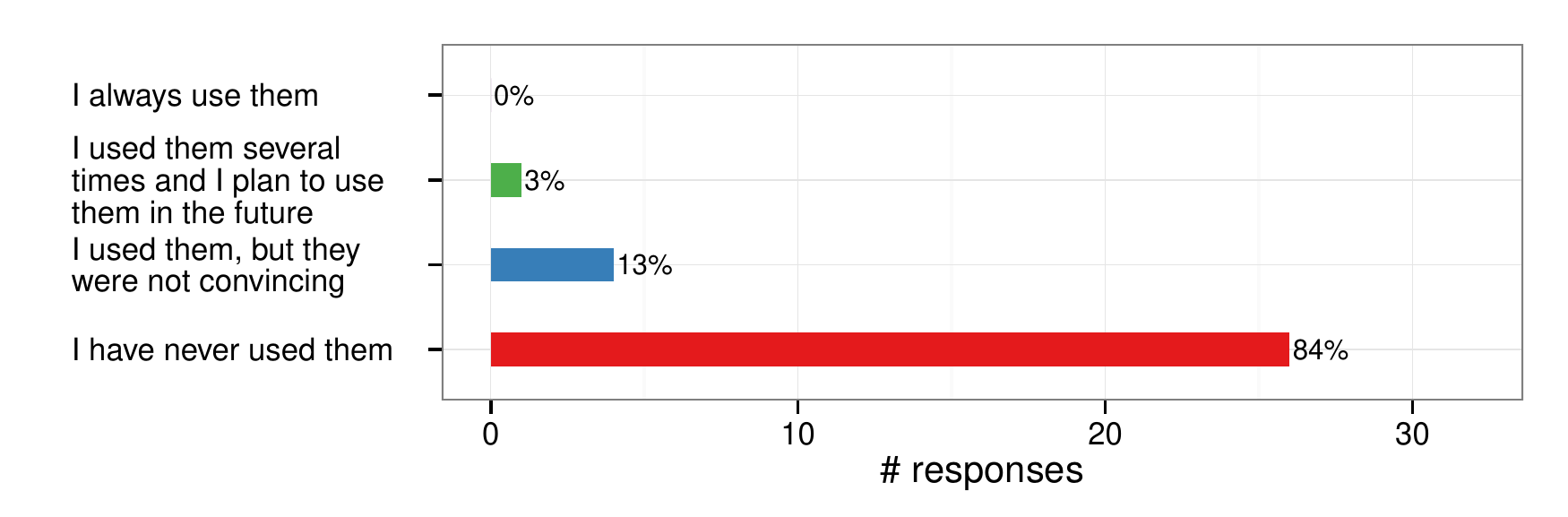} 

}

\end{knitrout}


\subsubsection{Do you know the differences between the available common open source licenses?}

\begin{knitrout}
\definecolor{shadecolor}{rgb}{0.969, 0.969, 0.969}\color{fgcolor}

{\centering \includegraphics[width=.8\linewidth]{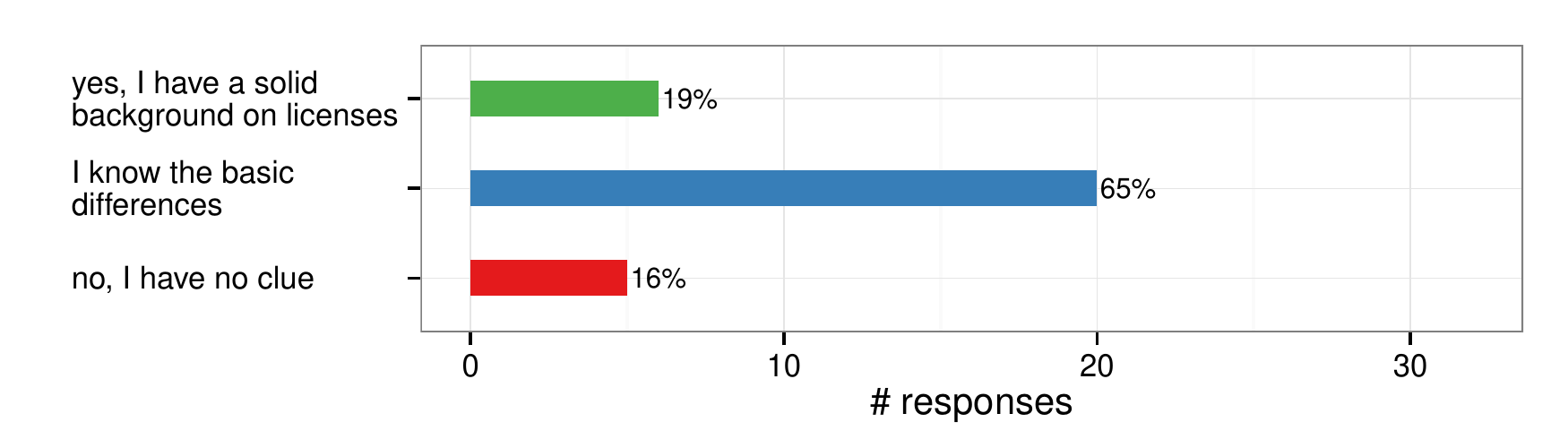} 

}

\end{knitrout}


\subsubsection{Do you know the policy used by YOUR research institution concerning the choice of open source licenses?}

\begin{knitrout}
\definecolor{shadecolor}{rgb}{0.969, 0.969, 0.969}\color{fgcolor}

{\centering \includegraphics[width=.8\linewidth]{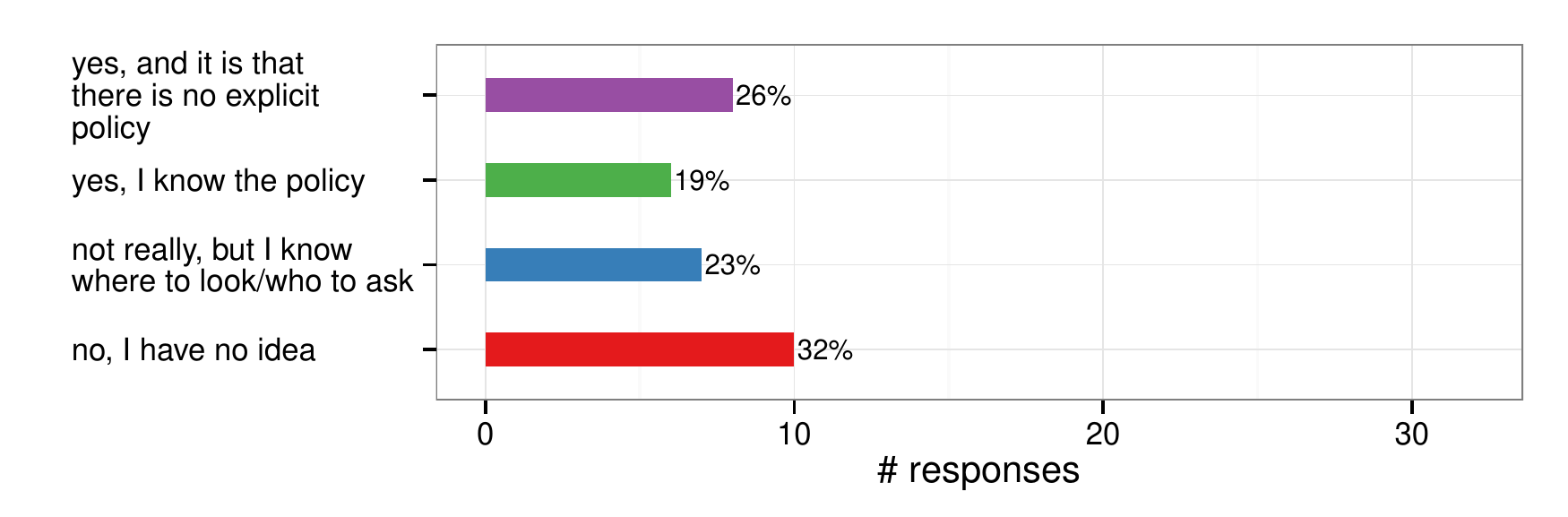} 

}

\end{knitrout}

\section{Conclusions}
\label{sec:conclusions}

We presented the poll results of a survey on reproducible research,
which had been conducted during the \europar conference 2015. Despite
the fact that only 31 persons completed the survey, the results give
us some evidence that reproducibility is a problem in our domain. In
fact, the survey revealed that the majority of the voters believe that
the state of reproducibility needs to be improved in the domain of
parallel and high performance computing. The survey participants also
think that the majority of the results presented in papers that they
receive for review are unlikely to be reproducible. The survey also
showed that scientists need to be better informed what the different
open-source licenses actually mean and which licenses are allowed to
be applied by their research institutions.  Last, we found evidence
that many scientists are not familiar with software for literate
programming and with scientific workflows, which can potentially help
to improve reproducibility of articles.

\bibliographystyle{abbrv}
\bibliography{reppar_survey}

\newpage

\appendix

\section{Original Questionnaire}
\label{sec:questionnaire}

\subsection{General Questions on Reproducibility}
\label{sec-1}

\begin{enumerate}
\item Do you care (in general) about the reproducibility of scientific
results (your own, others)?
\begin{enumerate}
\item no
\item yes
\end{enumerate}

\item Do you think you know what people mean when speaking about
"reproducible" results?
\begin{enumerate}
\item no
\item not sure, but I guess so
\item sure, I know what that means
\end{enumerate}

\item Do you know the difference between replicability, repeatability, and
reproducibility?
\begin{enumerate}
\item no
\item I am not sure, but I guess so
\item sure, I know what the differences are
\end{enumerate}

\item Do you think that the reproduction of already published results is
worth another publication?
\begin{enumerate}
\item yes, the reproduction alone is worth another publication
\item yes, but only if the publication contains new insights
\item no
\end{enumerate}

\item Have you tried to reproduce the results of others?
\begin{enumerate}
\item no
\item I tried once or twice
\item a couple of times (> 2 and <=10)
\item many times (>10)
\end{enumerate}
\end{enumerate}

\subsection{Current State of Reproducibility in Parallel Computing}
\label{sec-2}

\begin{enumerate}
\item Do you think the state of reproducibility for articles in our
research domain (Parallel Computing/HPC) needs to be improved?
\begin{enumerate}
\item no
\item yes
\end{enumerate}

\item Do you think current research articles in the domain of parallel
computing/HPC are reproducible by other independent researchers?
\begin{enumerate}
\item yes, all of them
\item yes, except a few papers (90\% reproducibility)
\item 50/50, some are, some are not
\item no, but a few might be reproducible (10\% reproducibility)
\item no article is reproducible (<1\% reproducible)
\item I really do not know
\end{enumerate}

\item Do you think that results published in top conferences (e.g.,
PPoPP, IPDPS) are generally easier to reproduce than those
published in lower-tier conferences in parallel computing (in the
last 5 years)?
\begin{enumerate}
\item yes, I know from my experience
\item probably, I can imagine that
\item I am note sure, but I guess not
\item no, not at all (all equally reproducible or non-reproducible)
\end{enumerate}

\item How often do you question the reproducibility of results when you
review other scientific articles?
\begin{enumerate}
\item for more than 90\% of the articles
\item for more than 50\% of the articles
\item for more than 10\% of the articles
\item never
\end{enumerate}
\end{enumerate}

\subsection{Reproducibility of Your Articles}
\label{sec-3}

\begin{enumerate}
\item Do you think the results (contribution) published in YOUR papers
are reproducible by others (in the last 5 years)?
\begin{enumerate}
\item yes, 100\%
\item most of them should be (>50\%)
\item I am not sure, I guess most results will not be (<= 50\%
reproducibility)
\item honestly, I know that they are not! (<5\% reproducibility)
\end{enumerate}

\item How often have you published the source code along with YOUR paper
(in the last 5 years)?
\begin{enumerate}
\item never
\item very few times (<25\%)
\item >= 25 \% and < 50\%
\item >= 50 \%
\item 100\% (for each article)
\end{enumerate}

\item Do you consider freely accessible HPC systems a necessity for
getting reproducible performance figures?
\begin{enumerate}
\item yes, for all studies
\item yes, but only for some studies
\item no
\end{enumerate}

\item How do you provide "source code" for others? 
\begin{enumerate}
\item I do not provide the source code
\item as an email attachment in response to a direct request
\item as an archive (zip, tar) on a personal webpage
\item I use public revision control system (e.g., GitHub)
\end{enumerate}

\item How do you provide the "raw data (of experiments)" for others?
\begin{enumerate}
\item I do not provide the raw data
\item as an email attachment in response to a direct request
\item as an archive (zip, tar) on a personal webpage
\item I use public revision control system (e.g., GitHub)
\end{enumerate}

\item How do you provide the "data analysis procedure (R scripts, etc)" for others?
\begin{enumerate}
\item I do not provide the data analysis procedure
\item as an email attachment in response to a direct request
\item as an archive (zip, tar) on a personal webpage
\item I use public revision control system (e.g., GitHub)
\end{enumerate}

\item How do you document how to use your source code / data analysis
scripts for others?
\begin{enumerate}
\item I do not document them
\item simple README files
\item standard documentation system (e.g., doxygen)
\item electronic laboratory notebook
\end{enumerate}

\item What are the main reasons for NOT making the source code/raw
data/data analysis procedure available? (multiple options)
\begin{enumerate}
\item it does not apply to me (as I make them available)
\item Technical difficulties. Lack of suited tools or hosting
infrastructure
\item Institution policy or legal aspects
\item I want to retain a competitive advantage
\item it is too time consuming
\item it is not rewarding
\item it is irrelevant because evolution is too fast
\item other
\end{enumerate}

\item What do you think will be the main difficulties/obstacles when
other independent researchers try to reproduce your experiments?
(multiple options)
\begin{enumerate}
\item the lack of access to specific machines
\item the lack of a specific software setup
\item the lack of documentation
\item the lack of time to reproduce our results
\item the lack of scientific credits (others will not get many
credits for reproducing our results)
\item other
\end{enumerate}
\end{enumerate}

\subsection{Tools/Software/Licenses for Reproducible Research}
\label{sec-4}

\begin{enumerate}
\item Have you used statistical software packages (e.g., R, SAS, SPSS,
..) for analyzing your experimental results?
\begin{enumerate}
\item no, not at all
\item not on a regular basis
\item yes, I always use them
\end{enumerate}

\item How would you rate YOUR knowledge of the programming language "R"?
\begin{enumerate}
\item I have never heard of R
\item I am a novice
\item I can code if needed, but I would not call myself an expert
\item I am an advanced user (expert)
\end{enumerate}

\item Do you use/have you used tools for literate programming (e.g., knitr,
org-mode, ..) for publishing articles?
\begin{enumerate}
\item never
\item I used them, but was not convinced
\item I used them and will do so in the future
\item I always use them
\end{enumerate}

\item Do you have practical experiences with workflow tools to support
reproducible research (e.g., VisTrails, Kepler, DataMill, etc.)?
\begin{enumerate}
\item I never used them
\item I used them, but they were not convincing
\item I used them several times and I plan to use them in the future
\item I now use them all the time
\end{enumerate}

\item Do you know the differences between the available common open source
licenses?
\begin{enumerate}
\item no, I have no clue
\item I know the basic differences
\item yes, I have a solid background on licenses
\end{enumerate}

\item Do you know the policy used by YOUR research institution
concerning the choice of open source licenses?
\begin{enumerate}
\item I have no idea
\item not really, but I know where to look/who to ask
\item yes, I know the policy
\item yes, and it is that there is no explicit policy
\end{enumerate}
\end{enumerate}



\end{document}